# A comparative study of the lipid panel levels at different duration time and temperature storage

Mohammed Mahdi Sami[1a] Mataz J. Jamai[1b] Bushra A. Jassam[2c]

[1]Al-Karkh University/college of remote sensing and geophysics [2]Al-Hikma University college/medical laboratory techniques
[a]mohamedmahdi81@kus.edu.iq, [b]mutaz.alwesy@kus.edu.iq [c]bushra.jassam@hiuc.edu.iq

**Abstract**

**Background:** the stability of serum specimen during time storage is importance in clinical and medical science researches in addition of diagnosis. Lipids are organic molecules that classified into 8 classes: fatty acids, phospholipids, glycerolipids, saccharolipids, polyketides, prenol lipids, sterol, and sphingolipids. Lipid analysis (cholesterol, triglyceride, high density lipoprotein, low density lipoprotein, very low-density lipoprotein) is a vital tool for diagnosis of numerous disease such as cardiovascular disease. follow-up the lipid profile progression is essential of numerous diseases.
**Methods:** a 72 apparently healthy individual were participated in current study the serum sample was taken after 12-14 hours of fasting. The serum cholesterol, triglyceride, high density cholesterol, low density cholesterol, and very low-density cholesterol were determined.
**Results:** there are no significant value found in all groups when compared to control except triglyceride that was significant after 24 hours of freezing.
**Keyword:** storage, temperature, cholesterol, triglyceride, HDL, LDL, and VLDL

## Introduction

It is important to have knowledge of which serum parameter affected during storage depending on time and temperature, in addition of other sample collecting techniques to increase the precision of results, whether it is research, diagnosis or follow-up. The stability of serum specimen during time storage is importance in clinical and medical science researches in plus of clinical diagnosis. Also, the duration of sample storage depends of which test want to be performed at which time. Some biochemical parameters such as acids, hormones, and enzyme need to performed immediately after specimen collection while other are stable for more than a week by freezing. So, knowing the stability of the parameters is an important and indispensable feature for obtaining precise results. In addition of procedure that should be use to avoid sampling error, the temperature, and duration of storage are essential factors that must be included in order to sustain the composition and integrity of specimen over the sample collection and assure accuracy of analytical results. Bonini and his coworker, Cuhadar and his colleagues mention that over75% of the errors in performing the final results occur in the pre-analytical phase (Bonini *et al.,* 2002; Cuhadar *et al.,* 2013). Lipids are organic hydrophobic molecules playing a vital role in cell signaling, cytoplasmic membranes, and metabolism Fahy E., *et al,* 2005. They classified into 8 classes: fatty acids, phospholipids, glycerolipids, saccharolipids, polyketides, prenol lipids, sterol, and sphingolipids (Fahy E., *et al* 2009). Plasma or serum reflects the lipid profile status, that use as biomarker to describe pathophysiological activity and processes (Ravi K., 2015). Its also, use to help in designing, synthesis and/or discover a new and more affective and sensitive drugs (Desai N., *et al,* 2013).

Lipid analysis is a vital tool for diagnosis of numerous disease such as cardiovascular disease (Vasile V., and Jaffe A., 2014). follow-up the lipid profile progression is essential of numerous disease such as hypertension, cardiovascular disease, and diabetes (Chopra I., and Kamal K., 2014;

Vasile V., and Jaffe A., 2014). It is important to get knowledge for collecting of sample, storage, handling and time of assay whether for diagnosis or research to reach a precise result (Müller H., 2021).

To determination the lipid profile [triglyceride (TG), cholesterol, high density lipoprotein (HDL), low density lipoprotein (LDL), and very low-density lipoprotein (VLDL)] the stability of sample is the most important criteria during pre-analysis phase. The stability can be described as a capacity to keeping the concentration of the analyte at minimal condition that may affect the results during a period of time (Simo *et al*, 2001; Evans *et al*, 1997).

Today there is no fix or universal procedures and information about the condition of storage and handling that act to minimize the concentration of analyte (Pahwa M., *et al*, 2015). There are few studies that examine the effect of storage and handling of specimen on the concentration of TG and cholesterol. The results of these studies were varying, some of them mention that storage time and freezing rise the concentrations of lipids (Maduka I., *et al*, 2009; Evans *et al*, 1997; Evans K., and Laker M., 1995; Pini *et al*, 1990; Katan and Tiedink 1989, while other show that, the levels of lipid concentrations are decrease at different duration time (Simo *et al*, 2001; Ekbom *et al*, 1996; Evans *et al*, 1997; Donnelly *et al*, 1995; Bausserman *et al*, 1994; Nanjee and Miller 1990; Tiedink and Katan 1989); and other mention the varies results (wood B., *et al*, 1982,1980 ).

There are studies show no significancy found in the levels of lipids concentration (Bausserman *et al*, 1994; Donnelly *et al*, 1995; Nanjee and Miller 1990; Stokes *et al*, 1986; Kuchmak et al. 1982; Wood *et al*, 1980).
Evans and his coworker Evans K, *et al*, 1997) investigate the effect of storage at −70°C on the levels of lipids. They mentioned that cholesterol is stable for about six months, while there was an increase in the concentration of TG. Other study shows that (Stokes, *et al*, 1986) no significant differences was found in the levels of concentrations of TG after 4 months of storage at −15°C.

The aim of current study is to investigate the levels of lipid profile and lipoprotein lipase activity at different incubation temperature and short period of storage time.

**Materials and methods**

A total of 72 apparently healthy individuals were enrolled in present study include (43 male and 29 female) age ranged 35-56. A blood was obtained from volunteers after 12-14 hours of fasting. The lipid panel include cholesterol, triglyceride, HDL, LDL, and VLDL were performed by using commercial kit manufactured by Biosystem. The serum sample was divided into four groups, as fresh sample or control (A) in which lipid panel was performed immediately after sample collecting, incubated at 37 °C for five hours (B), incubated at 8 °C for five hours (C), and freeze at -4 °C for 24 hours (D) until time assay. The data was analyzed using SPSS paired test as Mean ± standard deviation (SD), *P*-value considered a non-significant if it is more than 0.05 is, less than 0.05 is considered a significant and highly significant when *p*-value is less than 0.01.

**Results and discussion**

The concentration of the lipid panel (cholesterol, triglyceride, HDL, LDL and VLDL) were determined in volunteers with mean±SD age 43±6.92 and listed in Table 1.

Table (1): Mean±SD of cholesterol, triglyceride, HDL, LDL, and VLDL of A, B, C, and D groups.

| Parameters | A | B | C | D |
|---|---|---|---|---|
| Cholesterol | 170.19±21.9 | 169.37±26.03 | 177.48±19.1 | 171.42±22.8 |
| Triglyceride | 143.67±23.62 | 141.19±19.71 | 145.22±20.60 | 131.13±26.86* |
| HDL | 32.20±6.07 | 34.27±6.36 | 37.43±11.11 | 33.65±16.94 |
| LDL | 99.25±25.04 | 96.86±26.00 | 98.11±21.1 | 90.43±23.1 |
| VLDL | 28.74±4.62 | 28.24±6.35 | 29.05±4.03 | 24.43±5.25 |

The results indicant non-significant $p>0.05$ differences were found in the levels of serum cholesterol in all study groups Table 1. Most of studies mention the stability of serum cholesterol after storage under -4°C Ugwuezumba P., *et al*, 2018; Maguta I., *et al*, 2009. The concentration of serum triglyceride was determined and the results indicate non-significant $p>0.05$ decrease found in group B and C when compared to the control, while it was significantly decreased (about 8.6%) in group D Table 1. The decreasing of triglyceride concentration noted in previous studies, in which Ugwuezumba and his coworker found the level of triglyceride was decreased about 20.1% of its initial concentration after 7 days of storage at -70°C Ugwuenzumba *et al*, 2018, other study mentioned decreasing of serum triglyceride at the same condition and it was recorded about 38% decreased from its initial concentration Samuel Y., and Turkington V., 1975. Otherwise, there is studies mention no change in the level of the triglyceride after storage, Evan *et al*, 1995, Florian k., *et al*, 1994. These differences in the both groups of results may be due to varying in duration and temperature of storage in addition the methods of assay (enzymatic or alkaline hydrolysis) used, that can affect the results and quality of assay. The main reason for decreasing the level of triglyceride is due to that the activity of lipoprotein lipase (LPL) decrease after freezing and thawing (Margaret C., *et al*, 1991). The LPL is responsible to cleavage the ester bond in triglyceride to produce fatty acid (FA) and glycerol and the last is the essential as precursor molecule to react with ATP to produce glycerol-3-phosphate. The glycerol-3-phosphate oxidize to produce dihydroacetonephosphate and $H_2O_2$. The $H_2O_2$ interacts with 4-cholphenol to produce colored quinoneimine. The concentration of pink color quinoneimine is correspond to the level of triglyceride (Eggstein M., and Kreutz F., 1996 and stein E., 1987). The reason above can be supported by our results that the level of triglyceride was remain unchanged for 5 hours at 37 °C, at optimum activity of lipoprotein lipase that act to cleave all triglyceride molecules to glycerol and fatty acid. Serum HDL was showed non-significant increase in group A B, and C. these results was in line with Ugwuenzumba *et al*, 2018 and Sandra *et al*, 2017 how found non-significant increase in the level of HDL. The level of LDL and VLDL show non-significant in all studied group Table 1. This result was agreed with Maduka I., *et al*, 2009 and Sarah C., *et al*, 2003 who found 5.2% increase in the level of HDL when compared to control Table 1.

The following figure express the concentrations of lipid panel during different time of storage and temperature.

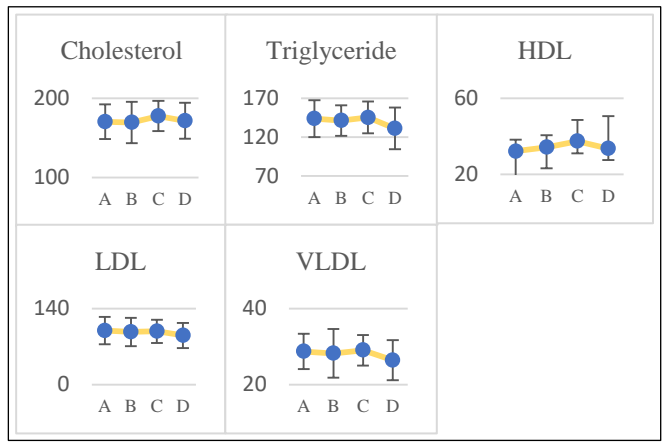

Figure (1): Mean±SD of cholesterol, triglyceride, HDL, LDL, and VLDL graph of A, B, C, and D groups.

## Conclusion

This study revealed a statistically non-significant difference in the concentration of most lipid analytes for conditions under study as following:

1- The levels of cholesterol are likely stable at different duration of storage and degrees.
2- The levels of triglyceride are showed stability after 5 hours incubation at 6°C and 36°C.
3- The concentration of triglyceride was significant decreased after 24 hours of storage at freezing state.
4- The levels of HDL, LDL, and VLDL were stable with storage at condition under study.

**Acknowledgements**
The authors acknowledge the volunteers that participate the blood sample.
**Author contribution:**
All authors contributed equally in this paper
**Data availability statement:**
No data associated in the manuscript